\documentclass{mem}
\usepackage{natbib}\usepackage{txfonts}\usepackage{balance}
\usepackage{graphicx}
\usepackage[a4paper,breaklinks,dvipdfm]{hyperref}
\idline{}{1}
\begin{document}
\def\teff{$T\rm_{eff }$}
\def\kms{$\mathrm {km s}^{-1}$}

\title{
ALMA and CARMA observations of Brown Dwarfs disks: testing the models of dust evolution
}

   \subtitle{}

\author{
L. \,Ricci\inst{1}, 
L. Testi\inst{2,3},
A. Natta\inst{3,4},
A. Scholz\inst{4},
I. de Gregorio-Monsalvo\inst{2,5},
A. Isella\inst{1} \and
J. M. Carpenter\inst{1}
%\and B. \, Rabbit\inst{1,2}
          }

%  \offprints{\email{lricci@astro.caltech.edu}}

\institute{
Department of Astronomy -- California Institute of Technology, MC 249-17, Pasadena, CA 91125, USA \and
European Southern Observatory, Karl-Schwarzschild-Strasse 2, D-85748 Garching, Germany \and
INAF-Osservatorio Astrofisico di Arcetri, Largo E. Fermi 5, I-50125 Firenze, Italy 
\and
School of Cosmic Physics, Dublin Institute for Advanced Studies, 31 Fitzwilliam Place, Dublin 2, Ireland 
\and
Joint ALMA Observatory (JAO)/ESO. Alonso de Cordova 3107. Vitacura 763 0335. Santiago de Chile \\
\email{lricci@astro.caltech.edu}
}

\authorrunning{Ricci et al.}

\titlerunning{ALMA and CARMA observations of BD disks}

\abstract{The first steps toward planet formation involve the coagulation of small microscopic grains into larger and larger pebbles and rocks in gas-rich disks around young stars and brown dwarfs. Observations in the sub-millimeter can trace mm/cm-sized pebbles in the outer disks, and investigate the mechanisms of coagulation/fragmentation and radial migration of these solids. These represent key, yet not fully understood ingredients for our understanding of the formation of planetesimals, the building blocks of planets.   
Here we present the first results from an observational program using the ALMA and CARMA sub-mm/mm interferometers aimed at characterizing the dust properties and disk structure of young disks around brown dwarfs and very low mass stars. Given the physical conditions expected for these disks, they represent critical test beds for the models of the early stages of planet formation in proto-planetary disks. 

%\keywords{Stars: abundances --
%Stars: atmospheres -- Stars: Population II -- Galaxy: globular clusters -- 
%Galaxy: abundances -- Cosmology: observations }
}
\maketitle{}

\section{Introduction}

The formation of planets involves a huge growth of solids starting from sub-$\mu$m grains and proceeding all the way up to terrestrial, ice and giant planets, with sizes up to hundreds thousands kilometers. Nearly all of this solid growth occurs in disks surrounding young pre-Main Sequence (PMS) stars. This whole process involves a variety of different physical mechanisms, the importance of which strongly depends on the level of aerodynamical coupling between the solids of different sizes and the gas in the disk \citep[for a review see][]{Chiang:2010}.  

The first stages of planet formation comprise the growth of dust grains into $\sim$ km-sized bodies called ``planetesimals''. At the typical gas densities of young circumstellar disks, grains with sizes of the order of $\sim 1 - 100~\mu$m or smaller are very well coupled to the gas in the disk and basically follow the dynamics of gas. These small grains collide at relative velocities which are determined by the equations of Brownian motion, and turn out to be much lower than 1 m/s.  At these collision velocities small grains stick very efficiently, with binding forces of electromagnetic nature, i.e. van der Waals forces. 

The situation changes drastically when solids with sizes of the order of $\sim 1-10$ mm (pebbles) to $\sim 1$ m (rocks) are formed in the disk. 
These solids are much less efficient in sticking than smaller particles, making fragmentation a more likely outcome of the collisions between these bodies.   
Since the surface-to-mass ratio decreases for solids of larger sizes, these solids start to be less coupled to the gas, and their dynamics changes accordingly. Because of the pressure radial gradient in the disk, gas rotates around the central star with a sub-keplerian speed. Pebbles and rocks do not feel any pressure force, they rotate at keplerian speed and are therefore \textit{faster} than gas at the same location in the disk. These solids perceive gas as a headwind which makes them lose angular momentum and migrate radially toward the central star \citep[][]{Weidenschilling:1977}. 

These two effects, fragmentation and fast radial migration, are the main obstacles to our understanding of planetesimal formation. As planetesimals are considered as the building blocks for larger bodies, understanding the behavior of these solids in real disks is critical for the whole process of planet formation. 

The presence of mm/cm-sized pebbles in real disks can be inferred observationally by measuring the spectral index $\alpha$ of the sub-mm/mm spectral energy distribution (SED), $F_{\rm{sub-mm}} \propto \nu^{\alpha}$. For optically thin dust emission, the spectral index of the SED is directly related to the spectral index $\beta$ of the dust emissivity coefficient, $\kappa_{\nu} \propto \nu^{\beta}$, where the relation is as simple as $\beta = \alpha - 2$ if dust emission is in the Rayleigh-Jeans regime at the frequencies of interest.  
While grains with sizes $< 10 - 100~\mu$m are characterized by $\beta \approx 1.5 - 2.0$, pebbles with sizes of the order of 1~mm and larger have $\beta < 1$ \citep[e.g.][]{Natta:2007}.  Therefore, the analysis of the sub-mm SED provides a strong tool to investigate mm/cm-sized pebbles in young disks and can be used to test models of solids evolution \citep{Birnstiel:2010}. 

Here we present the first results from an observational program using the ALMA and CARMA sub-mm/mm interferometers aimed at characterizing the dust properties and disk structure of young disks around brown dwarfs (BDs) very low mass stars (VLMs). At the different physical conditions expected for these disks when compared with disks around more massive young Solar-like stars, both the fragmentation and rapid radial drift problems for the growth of solids are amplified \citep[see][]{Pinilla:2013}. For this reason disks around VLMs and BDs are critical test beds for models of dust evolution in proto-planetary disks.

%Standard Big Bang cosmology predicts that protons and neutrons are
%assembled into nuclei of D, $^4$He, $^3$He and $^7$Li during the first
%three minutes \citep{wagoner}. The abundances of these nuclei are 
%a function of the baryon to photon ratio, which is simply related
%to the baryonic density of the Universe $\Omega_b$. 
%The concordance, within errors, of the measured primordial abundances 
%for a unique value of $\Omega_b$
%has been considered for the last twenty years one of the 
%observational pillars which support Big Bang cosmology.
%The opportunity of measuring the primordial Li abundance arises
%from the observation that the Li abundance in warm halo 
%dwarf stars is constant, regardless of the metallicity
%and the effective temperature \citep{spite82}, the so-called
%``Spite plateau''.
%In order to intepret this constant Li value as the primordial  
%one two conditions must be fulfilled: a) the Li present in the
%stellar atmosphere must have been preserved throughout the stellar
%evolution ; b) the Li produced by Galactic sources must 
%be neglegible compared to that produced in the Big bang.

%%
\begin{figure}[t!]
%\resizebox{\hsize}{!}{\includegraphics[clip=true]{co_mom0.eps}}
\includegraphics[clip=true,scale=0.35,trim=0 0 0 70]{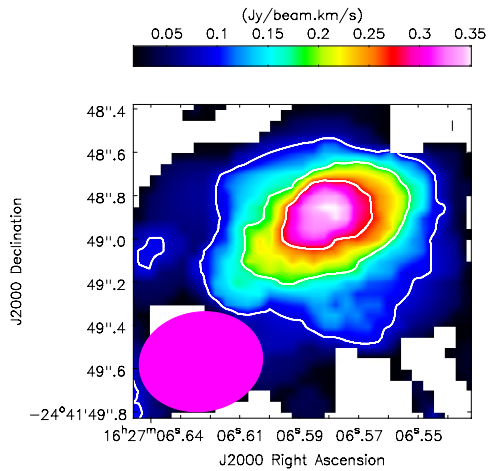}
\vspace{-1.3cm}
\caption{\footnotesize
ALMA moment 0 map of $\rho$-Oph 102 in CO($J = 3 - 2$). White contour lines are drawn at 2$\sigma$, 4$\sigma$, 6$\sigma$, where $\sigma = 45$ mJy/beam$\cdot$km/s is the measured rms noise on the map. The magenta color in the filled ellipse in the lower left corner indicates the size of the synthesized beam, i.e. FWHM $= 0.57'' \times 0.46''$, P.A. $=99^{\rm{o}}$ \citep[from][]{Ricci:2012b}.
}
\label{co}
\end{figure}

\section{ALMA and CARMA observations}

%
%\begin{table*}
%\caption{Abundances for TO stars in M 92}
%\label{abun}
%\begin{center}
%\begin{tabular}{lccccccc}
%\hline
%\\
%Star \# & [Fe/H] & $\sigma$ & [Mg/Fe] & [Ca/Fe] & [Ti/Fe] & [Cr/Fe] & [Ba/Fe] \\
%\hline
%\\
%18  &$ -2.63 $ & $0.22 $& $-0.02 $ &$+0.21$ & $+0.28$ & $-0.29 $ &$+0.11$ \\
%21  &$ -2.57 $ & $0.27 $& $-0.55 $ &$+0.17$ & $+0.44$ & $-0.22 $ &$-0.18$ \\
%34  &$ -2.58 $ & $0.24 $& $+0.06 $ &$+0.04$ & $+0.40$ & $   -  $ &$-0.05$ \\
%46  &$ -2.38 $ & $0.22 $& $-0.31 $ &$+0.22$ & $+0.17$ & $-0.27 $ &$-0.28$ \\
%60  &$ -2.54 $ & $0.30 $& $+0.12 $ &$+0.43$ & $+0.44$ & $-0.06 $ &$+0.39$ \\
%350 &$ -2.37 $ & $0.30 $& $-0.17 $ &$+0.37$ & $+0.34$ & $  -   $ &$  -  $ 
%\\
%\hline
%\end{tabular}
%\end{center}
%\end{table*}
%

\subsection{$\rho-$Oph 102} 

The first object that was observed under our ALMA program is $\rho-$Oph 102. This is a M5.5-M6 spectral type object with an estimated mass of $60~M_{\rm{Jup}}$ \citep{Natta:2002} in the Ophiuchus star forming region (SFR; age of $\sim 1$ Myr).
$\rho-$Oph 102 is known to be surrounded by a disk of dust from infrared observations, to have a significant mass accretion rate and to drive a wind and molecular outflow \citep[][]{Bontemps:2001,Natta:2002,Natta:2004,Whelan:2005,Phan-Bao:2008}.

We observed $\rho-$Oph 102 using ALMA Early Science in Cycle~0 at about 0.89~mm (Band~7) and 3.2~mm (Band 3). The ALMA Band~7 observations provided the first clear detection of molecular CO gas in a BD disk (Fig.~\ref{co}). By assuming optically thick gas emission and a temperature lower than $\sim 30 - 40$ K, a reasonable limit for a disk heated by the radiation from a BD with the properties of $\rho-$Oph 102, we obtained a lower limit of $\approx 15$ AU for the outer radius of the disk. Future ALMA observations with better sensitivity and angular resolution will allow to constrain the radial structure of this BD disk.

The continuum dust emission was clearly detected at both ALMA bands. We measured a spectral index of $\alpha_{\rm{0.89-3.2mm}} = 2.29 \pm 0.16$. Models of disks heated by the radiation from the central sub-stellar object show that the results of ALMA observations are best reproduced by emission from dust with a spectral index of the dust emissivity coefficient  $\beta \approx 0. 4 - 0.6$ \citep[for more details on the disk modeling, see][]{Ricci:2012b}. This implies that the dust emitting at the ALMA wavelengths has grown to sizes of at least $\sim 1$~mm, similarly to what is typically found in disks around young PMS stars.

\begin{figure}[t!]
\includegraphics[clip=true,scale=0.37,trim=0 0 0 150]{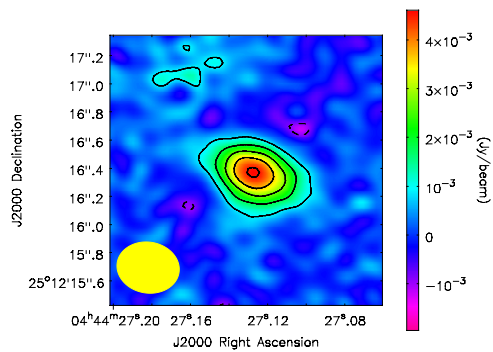}
\vspace{-2cm}
\caption{\footnotesize CARMA continuum map at 1.3 mm of the disk surrounding the young brown dwarf 2M0444$+$2512. Contours are drawn at $-3, 3, 6, 9, 12, 15\sigma$, where $1\sigma = 0.3$ mJy/beam is the rms noise. The yellow ellipse in the bottom left corner shows the synthesized beam, with FWHM $= 0.36'' \times 0.44''$, PA $= 80^{\rm{o}}$, obtained with natural weighting. 
However, the longest projected baselines of our observations probe angular scales as small as 0.16$''$ \citep[from][]{Ricci:2013}.
}
\label{2M0444_carma}
\end{figure}

\subsection{2M0444+2512}

2M0444$+$2512, hereafter 2M0444, is a young, $\sim 1$ Myr old, brown dwarf with a M7.25 spectral type in the Taurus star forming region. Like in the case of $\rho-$Oph 102, 2M0444, with  an estimated mass of about 50~$M_{\rm{Jup}}$ \citep{Luhman:2004}, is known to host a disk in dust from IR to sub-mm excess \citep{Knapp:2004, Scholz:2006, Guieu:2007, Bouy:2008}.

The disk around 2M0444 is brighter than the $\rho-$Oph 102 disk by a factor of $\sim 3$ in the sub-mm continuum, and we could detect its dust continuum emission at 1.3~mm using the CARMA interferometer \citep[][]{Ricci:2013}. With an angular resolution of 0.16$''$ corresponding to a physical scale of 22~AU (11~AU in radius) at the Taurus distance of 140 pc, the CARMA observations resolve the dust thermal emission from the disk (Fig. ~\ref{2M0444_carma}). By analyzing the interferometric visibilities using models of BD disks we obtained the first direct  constraints at sub-mm wavelengths for the radial distribution of dust in a BD disk.

\begin{figure*}[t!]
%\resizebox{\hsize}{!}{\includegraphics[clip=true]{co_mom0.eps}}
\includegraphics[clip=true,scale=0.77,trim=0 0 0 0]{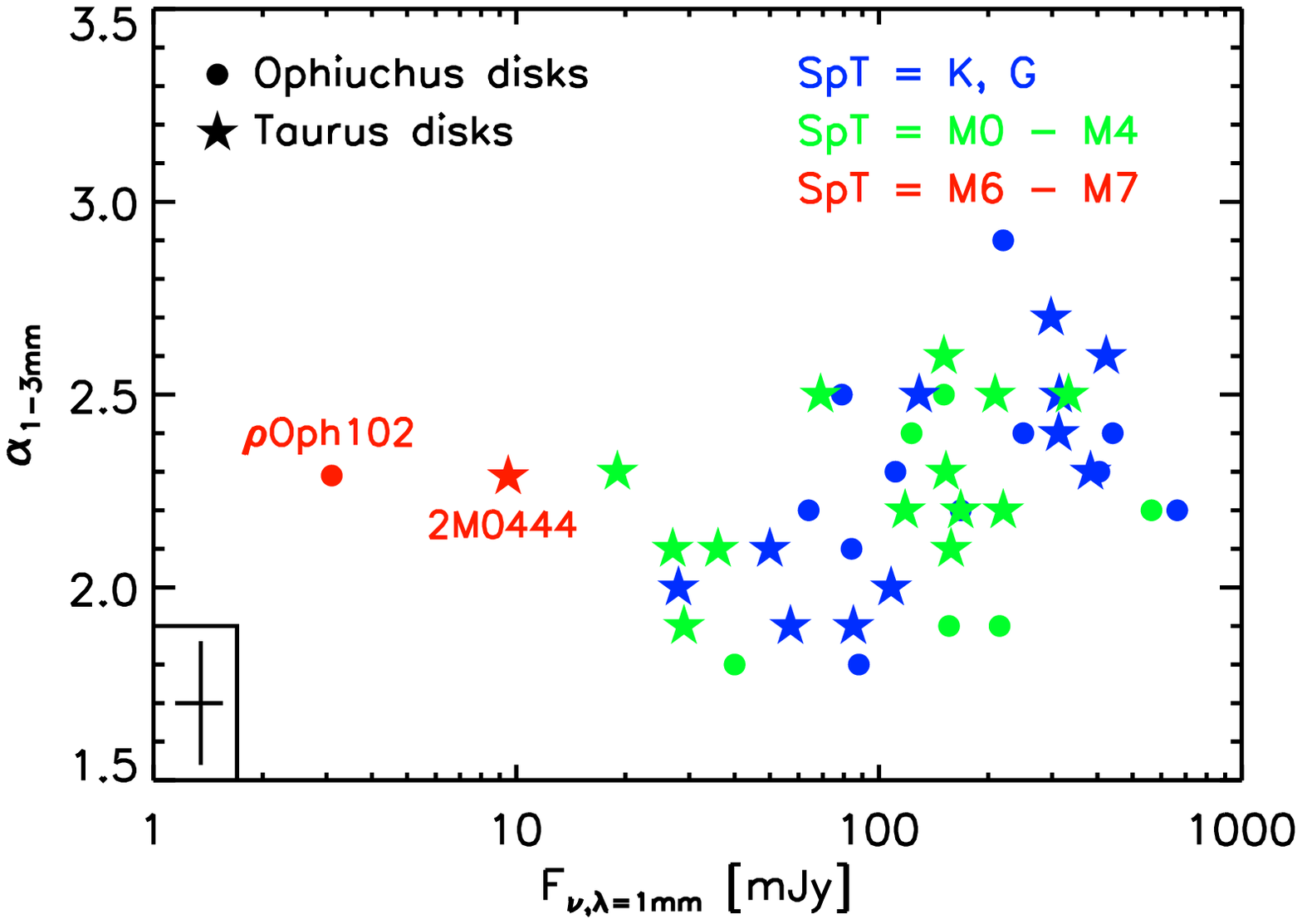}
%\vspace{-1cm}
\caption{\footnotesize Flux at 1 mm vs spectral index between 1 and 3 mm for disks around single PMS stars and brown dwarfs. Different colors and symbols refer to different stellar/sub-stellar spectral types and regions as indicated in the plot. Data for Taurus disks are from \citet{Ricci:2010a,Ricci:2012a}, for Ophiuchus disks from \citet{Ricci:2010b}, for 2M0444 from \citet{Bouy:2008}, \citet{Mohanty:2013} and \citet{Ricci:2013}, and for $\rho$-Oph 102 from \citet{Ricci:2012b}. Note that for same disks the values of the 1 mm flux density has been derived by interpolating between nearby wavelengths. The typical uncertainties of the data are shown in the lower left corner \citep[from][]{Ricci:2012b}.
}
\label{flux_alpha}
\end{figure*}

We measured a sub-mm spectral index of $2.30 \pm 0.25$ by combining the CARMA flux at 1.3~mm with the flux measurements by \citet{Mohanty:2013} and \citet{Bouy:2008} at 0.87~mm and 3.5~mm, respectively. This is similar to the value measured for the $\rho-$Oph 102 disk, and also for the disks surrounding more massive PMS stars (see Fig. \ref{flux_alpha}). Modeling of the visibilities rules out the possibility of a small ($R_{\rm{out}} < 10$ AU) optically thick disk. A dust emissivity spectral index $\beta = 0.50 \pm 0.25$ is needed to explain the sub-mm observations, thus finding evidence for grain growth to mm-sized pebbles also in the case of the 2M0444 BD disk.  

\section{Discussion and Conclusions}
We have presented the first results of a combined ALMA and CARMA observational program aimed at investigating the properties and radial distribution of dust and gas in disks around VLMs and BDs. These include the evidence for the presence of mm/cm-sized pebbles in the outer regions of two BD disks, $\rho-$Oph 102 and 2M0444, the detection of CO molecular gas in the $\rho-$Oph 102 disks, and direct constraints on the radial distribution of dust grains in the 2M0444 disk. 

Our ALMA Cycle 0 project contains two more disks in Taurus, other than 2M0444 which we also observed with CARMA. These ALMA data show values of the spectral index $\alpha$ which are very similar to the ones measured for $\rho-$Oph 102 and 2M0444, showing that mm-sized pebbles are regularly found in the outer regions of BDs/VLMs disks. In the case of 2M0444, CO($J=2-1$) emission is detected at several frequency channels, and this is consistent with gas emission from a disk in keplerian rotation (Ricci et al., in prep.).

The observations presented here show that solid growth to mm-sized pebbles and the retention of these particles in the outer regions are as efficient in disks around VLMs and BDs as in disks around more massive PMS stars. At a first look, these results challenge models of dust evolution in disks, as both these processes should be less efficient at the physical conditions of disks with relatively low mass/density around BDs and VLMs \citep[][]{Birnstiel:2010}.
Based on the results of our ALMA and CARMA observations, \citet{Pinilla:2013} have recently investigated the evolution of solids in disks around BDs and VLMs from a theoretical perspective. They calculated the time evolution of solids starting from sub-$\mu$m sizes and accounting for coagulation/fragmentation and radial migration in gas-rich disks with properties similar to the disks discussed here. They used these models to predict sub-mm fluxes for their simulated disks. They found that, in order to reproduce the observational data, stringent requirements have to be invoked for their disk models: relatively strong inhomogeneities in the pressure field of the gas in the disk to slow down the radial migration of mm-sized pebbles, small disk outer radii ($R_{\rm{out}} \approx 15 - 30$ AU), a moderate turbulent strength ($\alpha_{\rm{turb}} < 10^{-3}$), and average fragmentation velocities for ices of about 10~m/s.

Interestingly, recent ALMA observations of the young transitional disk IRS 48 have revealed a very pronounced azymuthal asymmetry in the distribution of mm-sized pebbles, these solids being strongly concentrated in a peanut-like shaped region in the disk \citep{vanderMarel:2013}.  These observations have been interpreted as evidence for the trapping mechanism of mm-sized particles invoked by the \citet{Pinilla:2013} models to explain the sub-mm observations of young disks \citep[see also][]{Armitage:2013}.
Future ALMA observations of disks around BDs and VLMs with better sensitivity and angular resolution will further test the models of the early evolution of solids in disks and shed more light to our understanding of the process of planet formation.
 
\begin{acknowledgements}
We would like to thank Antonio Magazzu, Eduardo Martin, and all other SOC and LOC members for the organization of this stimulating conference held at a fantastic venue.  This work makes use of the following ALMA data: ADS/JAO.ALMA$\#$2011.0.00259.S. ALMA is a partnership of ESO (representing its member states), NSF (USA) and NINS (Japan), together with NRC (Canada) and NSC and ASIAA (Taiwan), in cooperation with the Republic of Chile. The Joint ALMA Observatory is operated by ESO, AUI/NRAO and NAOJ. The National Radio Astronomy Observatory is a facility of the National Science Foundation operated under cooperative agreement by Associated Universities, Inc. Support for CARMA construction was derived from the Gordon and Betty Moore Foundation, Kenneth T. and Eileen L. Norris Foundation, James S. McDonnell Foundation, Associates of the California Institute of Technology, University of Chicago, states of California, Illinois, Maryland, and NSF. Ongoing CARMA development and operations are supported by NSF under a cooperative agreement, and by the CARMA partner universities. We acknowledge support from the Owens Valley Radio Observatory, which is supported by the NSF grant AST-1140063.
\end{acknowledgements}

\bibliographystyle{aa}

\end{document}